\documentclass[
prc,%
10pt,%
final,%
notitlepage,%
oneside,%
twocolumn,%
nobibnotes,%
nofootinbib,
superscriptaddress,%
floatfix]%
{revtex4}

\usepackage{color} 
\usepackage{amsfonts} 
\usepackage{amsbsy} 
\usepackage{mathrsfs}
\usepackage{graphicx}
\usepackage{comment}

\def\lsim{\mathrel{\rlap{ \lower4pt\hbox{\hskip-3pt$\sim$}}
    \raise1pt\hbox{$<$}}} 
\def\gsim{\mathrel{\rlap{ \lower4pt\hbox{\hskip-3pt$\sim$}}
    \raise1pt\hbox{$>$}}} 
\def\scr#1{\mbox{\scriptsize #1}}

\begin{document}

\title{Space-time regions 
of high baryon density and baryon stopping in heavy-ion collisions
}

\author{Yuri B. Ivanov}\thanks{e-mail: yivanov@theor.jinr.ru}
\affiliation{Bogoliubov Laboratory of Theoretical Physics, JINR Dubna,
  141980 Dubna, Russia} 
\affiliation{National Research Center
  "Kurchatov Institute", 123182 Moscow, Russia}

\begin{abstract}
Four-volumes ($V_4=$ spatial-3-volume$\times$lifetime) are calculated within the 
model of three-fluid dynamics (3FD) and  
compared with those of the Jet AA Microscopic Transport Model (JAM). 
The calculations are performed for central Au+Au collisions at energies $\sqrt{s_{NN}}=$  3 -- 19.6 GeV.
These $V_4$ indicate optimal collision-energy ranges for realizing macroscopic high baryon-density matter.
It is found that the 3FD four-volumes noticeably exceed those in the JAM, which  
indicates a stronger baryon stopping in the 3FD model as compared to that JAM.
It is argued that this difference in the baryon stopping correlates with stiffness of the equation of state 
implemented in these models.
Contrary to JAM, the four-volume, where a baryon density ($n_B$) exceeds 
three times the normal nuclear density ($n_0$), does not exhibit a maximum
as a function of $\sqrt{s_{NN}}$.  
It decreases monotonically with increasing $\sqrt{s_{NN}}$, remaining 
at a fairly macroscopic level (i.e. $V_4\gsim 5.5^4$ fm$^4$/c).
For higher baryon densities, $V_4$  exhibits  maxima in its dependence on $\sqrt{s_{NN}}$.
The optimal energy range for densities $n_B/n_0>$ 4 is located at $\sqrt{s_{NN}}=$ 3.2--8 GeV. 
Even for $n_B/n_0>$ 6, the four-volume remains quite macroscopic ($V_4\gsim  4^4$ fm$^4$/c) 
at $\sqrt{s_{NN}}=$ 4.5--9 GeV contrary to the JAM.  
%
\end{abstract}
\maketitle

\section{Introduction}

At present, much attention is paid to collisions of heavy ions at energies of $\sqrt{s_{NN}}=$ 3–-20 GeV, 
at which the highest baryon densities are achieved \cite{Lovato:2022vgq,Sorensen:2023zkk}. 
This interest is caused by the onset of transition into quark-gluon phase (QGP) in heavy-ions collisions that is expected 
at these energies. Besides, critical point of the QGP phase diagram is presumably achieved at these energies 
\cite{Du:2024wjm,Stephanov:2024xkn}. 
The renewed interest in the equation of state (EoS)  of nuclear matter at the high baryon densities attainable in heavy-ion collisions at these energies is also supported by increased efforts to constrain the EoS of neutron-rich matter, probed in studies of neutron stars and neutron star mergers \cite{Baym:2017whm,Burgio:2021vgk}.  
Theoretical information on the QGP phase diagram at high baryon densities is inaccessible from first-principle lattice QCD  because of the notorious sign problem \cite{Nagata:2021ugx}. 
Therefore, only various QCD-motivated effective models are used. These models predict very rich phase structure at 
high baryon densities \cite{Fukushima:2010bq}, 
which additionally inspire experimental studies of heavy-ion collisions at high baryon densities. 
This energy range is actively investigated in the Beam-Energy Scan (BES) program and STAR-FXT (fixed target) 
at RHIC (Brookhaven) and NA61/SHINE at
SPS (CERN), and is planned to be further studied worldwide \cite{Galatyuk:2019lcf} at
NICA, FAIR, HIAF, and J-PARC-HI facilities.

Analysis of available data on hadron yields within the statistical model has shown that maximal baryon densities 
at the freeze-out stage are reached at collision energies  6--9 GeV \cite{Randrup:2006nr,Randrup:2009ch}. 
However, the baryon densities deduced from this analysis
are not the maximal ones achieved in the heavy-ion reactions, because 
the freeze-out occurs at a very late stage, when the baryon matter is diluted by the expansion.
Dynamical models are required to estimate the maximal densities attained in heavy-ion collisions.

Various dynamical models of heavy-ion collisions are extensively used 
to simulate the heavy-ion collisions at high baryon densities. 
The diversity of these models ranges from microscopic transport models 
\cite{Li:1995pra,Bass:1998ca,Buss:2011mx,SMASH:2016zqf,Aichelin:2019tnk,Nara:1999dz,Nara:2023vrq}, 
hybrid models \cite{Petersen:2008dd,Batyuk:2016qmb,Kozhevnikova:2020bdb,Akamatsu:2018olk,Cimerman:2023hjw,Shen:2022oyg}, 
to hydrodynamic models \cite{Ivanov:2005yw,Ivanov:2013wha}. 
The above list is far from being complete. 
These models use different input data: cross sections, (mean-field) potentials, EoSs, \textit{etc.}, 
which can lead to different dynamic evolution \cite{Arsene:2006vf} while giving  
similar results for observables.

Previous model calculations \cite{Arsene:2006vf} predicted that the maximum
baryon density at the center of the collision system can exceed $8n_0$ for $\sqrt{s_{NN}}>$ 5 GeV,  
where $n_0$ is the normal nuclear density. 
However, these high values of the densities are not very meaningful because they are achieved in a
tiny volume and maintained for a tiny period of time. It
is desirable to evaluate the density in a reasonably large
volume which survives for a reasonably long time span. 
This is crucial for preserving observable signatures of the dense matter in the final state.
However, the space volumes and lifetimes are not Lorentz invariant quantities. 
Therefore, space-time volumes occupied by high-baryon-density matter defined as 
\begin{eqnarray}
\label{V4}
V_4 (\widetilde{n}_B) = \int d^4 x \;\Theta (n_B(x)-\widetilde{n}_B)  
\end{eqnarray}
were proposed in Ref. \cite{Friman:1997sv} and then also used in 
Ref. \cite{Ivanov:2005yw,Ivanov:2019gxm}. Here, 
$n_B(x)$ is the local proper (i.e. in the local rest frame) baryon density,  
$\widetilde{n}_B$ is a threshold baryon density. The integration in Eq. (\ref{V4}) runs over 
space-time regions, where $n_B(x)>\widetilde{n}_B$, the step function $\Theta (z)$ equals 1 for $z\geq 0$ and 0 for $z<0$. 
The $V_4 (\widetilde{n}_B)$ is an invariant measure of the
space--time region, where the $n_B$ exceeds the threshold value $n_B(x)>\widetilde{n}_B$.  
In Ref. \cite{Taya:2024zpv}, it was suggested to compare the dynamic evolution of nuclear collisions
within different models by analyzing these space-time four-volumes occupied by high-baryon-density matter.

In this paper, predictions of the model of three-fluid dynamics (3FD) \cite{Ivanov:2005yw,Ivanov:2013wha} 
for the four-volumes (\ref{V4}) are presented and compared with those of 
the Jet AA Microscopic Transport Model (JAM) 
\cite{Nara:1999dz,Nara:2023vrq} that were reported in Ref. \cite{Taya:2024zpv}.

\section{3FD simulations} 
  \label{3FD simulations}

The 3FD model is a further development of the multi-fluid approaches previously worked out  
by Los-Alamos group \cite{Amsden:1978zz,Clare:1986qj}, at 
the Kurchatov Institute \cite{Mishustin:1989nj,Mishustin:1991sp}, 
Frankfurt University \cite{Katscher:1993xs,Brachmann:1997bq} and
GSI \cite{Russkikh:1993ct}.
The 3FD model \cite{Ivanov:2005yw,Ivanov:2013wha} simulates nonequilibrium at
the early stage of nuclear collisions by means of two counterstreaming baryon-rich fluids, 
which is   minimal way to model the early-stage nonequilibrium.
Newly produced particles predominantly populate the midrapidity region. 
Therefore, they are attributed to a third (``fireball'') fluid.
These three fluids, i.e. the projectile (p), target (t), and fireball (f), 
are governed by conventional hydrodynamic equations coupled by 
friction terms in the right-hand sides of the Euler equations. The friction terms 
describe the energy--momentum exchange between the fluids. 
In the hadronic phase, the friction was estimated
in Ref. \cite{Sat90}, this friction is used in the 3FD simulations. 
The phenomenological friction in the QGP was fitted to reproduce the baryon stopping
at high collision energies within the deconfinement scenarios as it is described in 
Ref. \cite{Ivanov:2013wha} in detail.
The 3FD model describes a nuclear collision from the stage of the incident cold nuclei
approaching each other, to the final freeze-out 
\cite{Russkikh:2006aa,Ivanov:2008zi}.  

The physical input of the 3FD calculations is
described in Ref. \cite{Ivanov:2013wha}. The 3FD simulations are traditionally 
performed with three different EoSs: a
purely hadronic EoS \cite{Mishustin:1991sp} and two versions of the EoS
involving the deconfinement transition \cite{Toneev06}, i.e. a first-order phase transition (1PT) 
and a smooth crossover one. This paper mainly demonstrates results with crossover EoS 
as the most successful one in reproduction of various observables in a wide range of collision energies, 
see, e.g., Ref. \cite{Ivanov:2025vru} and references therein. Results for the 1PT are also demonstrated 
to illustrate dependence on the EoS.

\begin{figure}[!tbh]
\includegraphics[width=.47\textwidth]{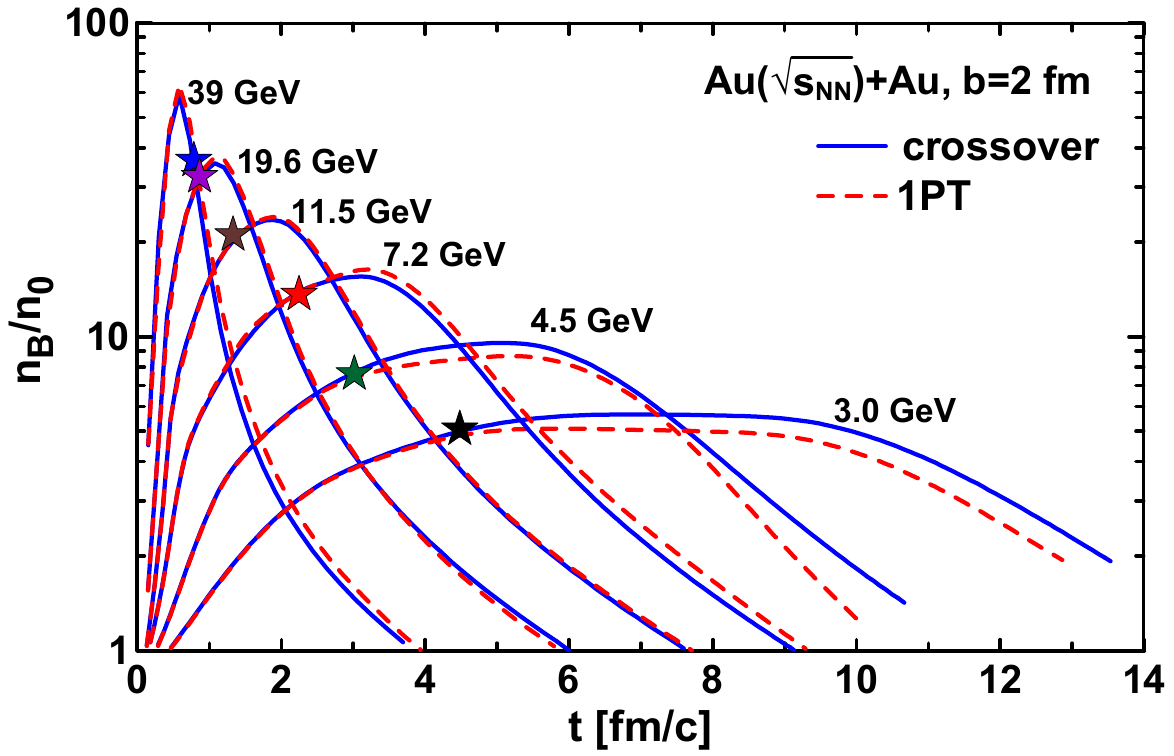}
  \caption{(Color online)
Time evolution of baryon density
in the central region of central ($b$ = 2 fm) Au+Au collisions at
various collision energies ($\sqrt{s_{NN}}$)
Results for  the crossover  and 1PT EoSs are displayed. 
Star symbols on the curves mark the time instant of the baryon-matter stopping.
}
\label{fig:nB_vs_t}    
\end{figure}

The counterstreaming of the p and t fluids takes place
only at the initial stage of the nuclear collision. At later
stages the baryon-rich (p and t) fluids have already either
partially passed though each other or partially stopped
and unified. The unification procedure is described 
in Ref. \cite{Ivanov:2005yw,Ivanov:2013wha}. 
The occurrence of such stopping is illustrated at the example of the time evolution of baryon density 
in the central region of central (i.e. at impact parameter $b$ = 2 fm) Au+Au collisions, see Fig. \ref{fig:nB_vs_t}. 
Star symbols on the curves mark the time instants of the baryon-matter stopping.
Details of this calculation can be found in Ref. \cite{Ivanov:2019gxm}.

As seen from Fig. \ref{fig:nB_vs_t}, the baryon stopping in the central region 
is achieved shortly before reaching the maximal baryon density at $\sqrt{s_{NN}}<$ 20 GeV. 
These maximal baryon densities are quite high.  However, these high values are reached only 
in a small spatial region of the colliding system and for a short time, very short time at the 
energy of  19.6 GeV, or not that short at lower collision energies. The four-volume of Eq. (\ref{V4})
provides a quantitative measure of the space-time region occupied by the dense baryonic matter. 
While Fig. \ref{fig:nB_vs_t} gives an idea of the lifespan of the compressed baryonic matter.

At collision energy of 39 GeV, the baryon stopping 
is achieved shortly \textit{after} reaching the maximal baryon density. 
This means that only a part of baryon charge is stopped even in the central region of central Au+Au collision. 
Another part passes though the counter-streaming baryon flow and forms dense baryonic blobs in the fragmentation regions.
Properties of these baryon-rich fragmentation regions (i.e. the baryonic fireballs)
produced in central  heavy-ion collisions have been discussed long ago 
\cite{Anishetty:1980zp,Csernai:1984qh,Gyulassy:1986fk,Frankfurt:2002js,Mishustin:2001ib}. 
Interest in this topic was revived \cite{Li:2016wzh,Li:2018ini,Ivanov:2017xee,Ivanov:2018rrb,Kolbe:2020hem}. 
after proposal \cite{Brodsky:2012vg}
to perform experiments at the Large Hadron Collider (LHC) at CERN in the
fixed-target mode  (AFTER@LHC experiment).

The ultra-relativistic energies are beyond the scope of the present paper. 
However, the important message from those energies is that the kinetically 
equilibrated baryonic matter may be produced not only in the baryon stopping but also 
as a result of passing the baryon-rich matter though the interaction region. 
Such kind of partial transparency takes place in peripheral regions of colliding Au+Au nuclei 
even at $\sqrt{s_{NN}}<$ 20 GeV.  Therefore, the term ``equilibrated baryonic matter'' is used 
for both stopped and passed-though baryonic matter.

The baryon stopping depends on the way of implementation of the interfluid friction in the model. 
The 3FD \cite{Ivanov:2005yw,Ivanov:2013wha} assumes that baryons always remain leading particles in elementary 
interactions, i.e. they always belong to the baryon-rich fluid (p or t) of their origin.  In terms 
of continuity equations for the baryon charge, it reads 
\begin{eqnarray}
\label{3FD}
\partial_\mu J^\mu_\alpha  = 0 \hspace{5mm} (\mbox{3FD}),  
\end{eqnarray}
where $J^\mu_\alpha$ is baryon current of $\alpha$ (= p,t, or f). 
Moreover, $J^\mu_{\scr f}\equiv 0$ because there is no baryon-charge transfer into f-fluid. 
The complete baryon stopping happens only due to unification of the baryon-rich fluid
\cite{Ivanov:2005yw,Ivanov:2013wha} when corresponding  conditions are met.  
In the recently developed next-generation hybrid three-fluid model for simulating heavy-ion collisions 
(MUFFIN\footnote{MUlti Fluid simulation for Fast IoN collisions}) \cite{Cimerman:2023hjw}, 
a refined “charge transfer” scheme \cite{Werthmann:2025ueu} of the interfluid friction was implemented 
\begin{eqnarray}
\label{MUFFIN}
\partial_\mu J^\mu_\alpha  = R_\alpha, \hspace{5mm} \sum_\alpha R_\alpha =0  \hspace{5mm} (\mbox{MUFFIN}),  
\end{eqnarray}
where $R_\alpha$ is a rate of the baryon-charge transfer from/to $\alpha$-fluid. 
This is a more flexible scheme of the baryon stopping, which however requires additional parameters. 
The population of the f-fluid by the baryon charge may lead to formation of a plateau 
in baryon rapidity distribution at the initial stage of the collision, 
which is important for description of the directed flow at high collision energies \cite{Du:2022yok}.

\section{Four-volume occupied by dense baryonic matter} 
  \label{Four-volume}

\begin{figure}[!hbt]
\includegraphics[width=6.1cm]{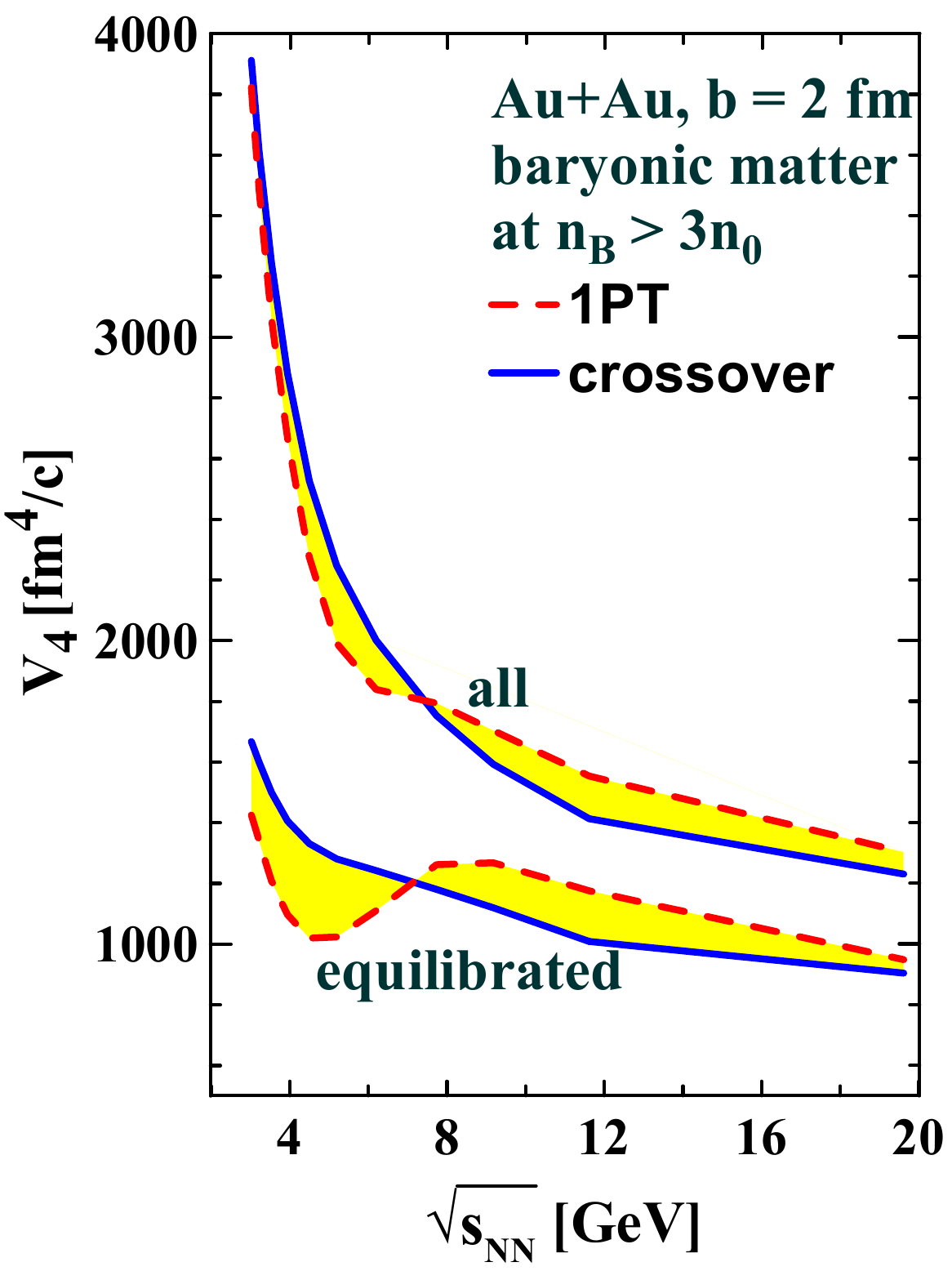}
 \caption{(Color online)
Four-volume, in which the baryon density exceeds value $3n_0$
in the central ($b$ = 2 fm) Au+Au collisions as function of collision energy
$\sqrt{s_{NN}}$.  Calculations are done with the crossover and 1PT EoSs. 
}
\label{fig:V4_sNN_3nB_lin}
\end{figure}

Four-volume, in which the baryon density exceeds value $3n_0$
($n_0 =$ 0.15 1/fm$^3$) 
in the central ($b$ = 2 fm) Au+Au collisions at various collision energies
$\sqrt{s_{NN}}$, is presented in Fig. \ref{fig:V4_sNN_3nB_lin}

The four-volume, occupied by all (i.e. not necessary equilibrated) baryon matter, is considerably larger 
than that of the equilibrated baryon matter, especially at lower collision energies. This is because the 
equilibration takes longer time at these lower collision energies.

The four-volume depends on the EoS. It is larger for softer EoS 
because a soft EoS is less resistant to compression of the matter. 
This similar to that in 
the JAM model \cite{Taya:2024zpv}: The cascade version of JAM (i.e. very soft EoS) predicts 
a $V_4$ larger than the mean-field version of JAM.   
At lower collision energies the system evolves within the hadronic sector of the 1PT EoS.  
The hadronic sector of the 1PT EoS is stiffer than the crossover EoS because the latter is softened 
by a small admixture of the QGP. Therefore,  $V_4^{\scr{crossover}}>V_4^{\scr{1PT}}$ at lower collision energies. 
At higher collision energies $\sqrt{s_{NN}}\gsim$ 7 GeV, the transition to QGP within the 1PT scenario has been already completed. 
Therefore, the 1PT EoS becomes softer than the crossover EoS, hence 
$V_4^{\scr{crossover}}>V_4^{\scr{1PT}}$ at higher collision energies. 
Thus, the wiggles in the 1PT curves in Fig. \ref{fig:V4_sNN_3nB_lin} reflect onset of the first-order phase transition.  
The onset of the phase transition within the 1PT EoS indeed occurs at these collision energies, 
as it is indicated by the proton directed flow  \cite{Ivanov:2025vru}.

\begin{figure}[!hbt]
\includegraphics[width=6.1cm]{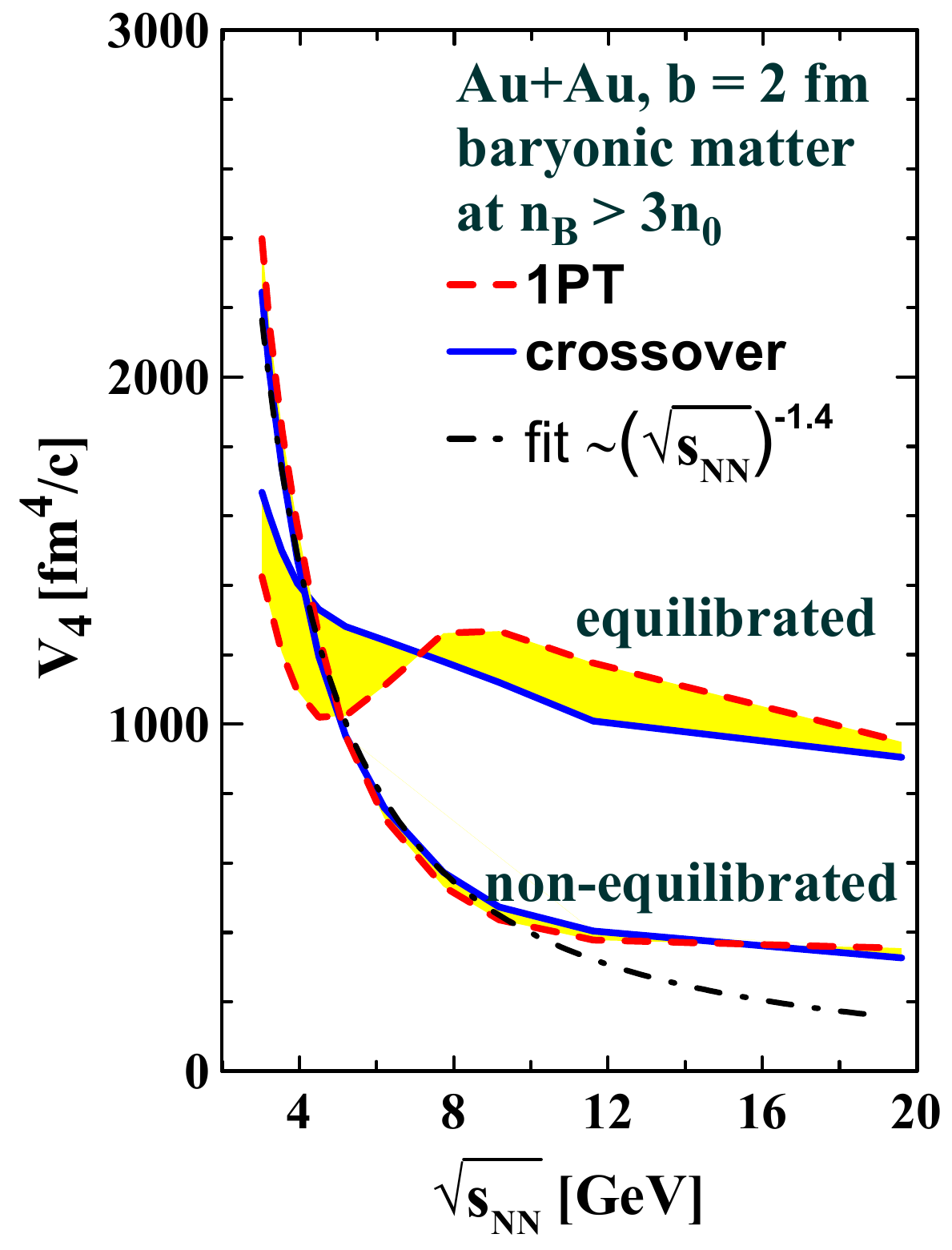}
 \caption{(Color online)
The same as in Fig. \ref{fig:V4_sNN_3nB_lin} but separately for equilibrated and nonequilibrated baryon densities. 
}
\label{fig:V4_noneq_sNN_3nB_lin}
\end{figure}

Figure \ref{fig:V4_noneq_sNN_3nB_lin} demonstrates separate contribution of the nonequilibrated baryon matter 
into the four-volume,   
$V_4(\widetilde{n}_B^{\scr{nonequilibrated}}) = V_4(\widetilde{n}_B^{\scr{all}}) - 
V_4(\widetilde{n}_B^{\scr{equilibrated}})$, 
in comparison with $V_4(\widetilde{n}_B^{\scr{equilibrated}})$. 
The $V_4(\widetilde{n}_B^{\scr{nonequilibrated}})$ is practically independent of the EoS. 
This means that the main contribution to $V_4(\widetilde{n}_B^{\scr{nonequilibrated}})$
comes from the interpenetration process of colliding matter without its real compression, 
thus that even small admixture of the QGP in the crossover EoS does not affect the matter. 
A real compression starts only after equilibration of the matter. 

The interpenetrating matter gives large contribution into $V_4$ because of simple kinematic factor. 
For instance, if two counter-streaming flows of normal nuclear density, $n_0$, 
interpenetrate each other with relative velocity of the initial colliding beams%
\footnote{This is a free-streaming regime in terms of Ref. \cite{Taya:2024zpv}}, 
then their baryon 
density in the their common rest frame is $2n_0\gamma$, where $\gamma=\sqrt{s_{NN}}/(2m_N)$ is the $\gamma$
factor of colliding beams, $m_N$ is the nucleon mass. If $\gamma>1.5$ 
(and it really is already at $\sqrt{s_{NN}}>$ 3 GeV), 
this interpenetrating matter gives contribution into $V_4(3n_0)$. 

In the 3FD model, the initial nuclei are presented by sharp-edge spheres of nuclear matter of normal nuclear density. 
This sharp-edge (rather than diffuse-edge) form is caused by reasons of numerical stability of the code. 
The sharp-edge approximation increases the free-streaming (FS)  contribution to $V_4(3n_0)$ as compared with that in 
a more realistic diffuse-edge prescription, as in the JAM. Indeed, when the aforementioned density $2n_0\gamma$
within the sharp-edge approximation contributes to $V_4(3n_0)$, the diffuse-edge density $2n_B\gamma$ (with $n_B<n_0$) 
does not.

Not all $V_4(\widetilde{n}_B^{\scr{nonequilibrated}})$ is due to FS contribution. 
This is seen from the energy dependence of $V_4(\widetilde{n}_B^{\scr{nonequilibrated}})$. 
If it were only the FS contribution, the energy dependence of $V_4(\widetilde{n}_B^{\scr{nonequilibrated}})$ 
would be%
\footnote{Of course, this is only in the sharp-edge approximation. In Ref. \cite{Taya:2024zpv} it is not so.}
$$
V_4(\widetilde{n}_B^{\scr{FS}}) = V_3 \Delta t \sim 1/\sqrt{s_{NN}(s_{NN}-4m_N^2)}, 
$$
because the Lorentz contracted spacial volume is $V_3\sim 1/\sqrt{s_{NN}}$, and the passage time of 
Lorentz contracted nuclei through each other is $\Delta t \sim 1/\sqrt{s_{NN}-4m_N^2}$. 
In fact, see fit of $V_4(\widetilde{n}_B^{\scr{nonequilibrated}})$ in Fig. \ref{fig:V4_noneq_sNN_3nB_lin}, 
the energy dependence is weaker at lower collision energies, $\sim [\sqrt{s_{NN}}]^{-1.4}$, 
which means that deceleration of the counter-streaming flows takes place. 
The deceleration increases the passage time.

Long equilibration time at lower $\sqrt{s_{NN}}$, as seen from Fig. \ref{fig:nB_vs_t}, leads to 
sharp increase of $V_4(\widetilde{n}_B^{\scr{nonequilibrated}})$ with decrease of the collision energy.

\begin{figure}[!t]
\includegraphics[width=6.1cm]{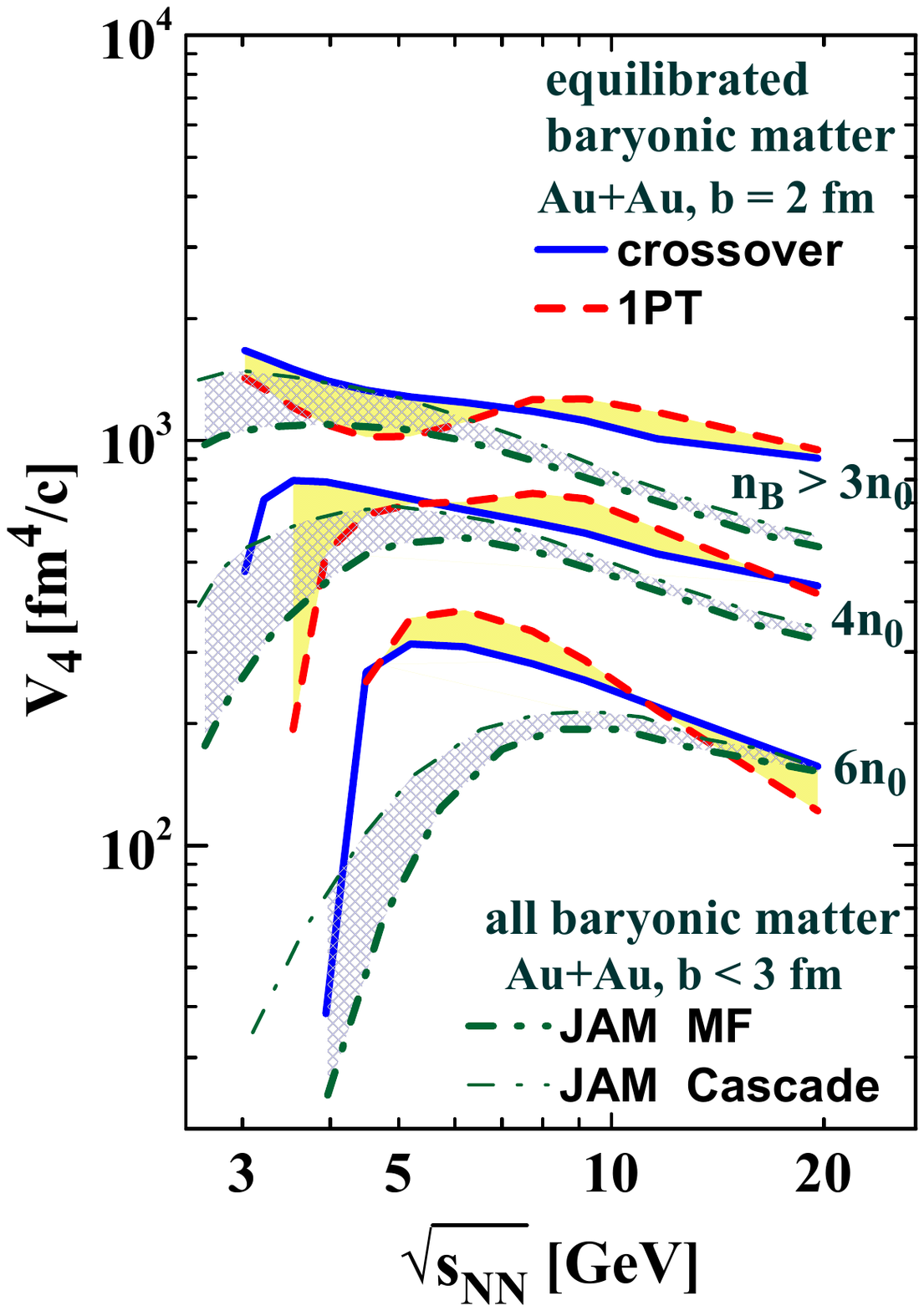}
 \caption{(Color online)
Four-volume, in which the baryon density of the equilibrated matter exceeds value $\widetilde{n}_B$
in the central ($b$ = 2 fm) Au+Au collisions, as function of collision energy
$\sqrt{s_{NN}}$.  Calculations are done with the crossover and 1PT EoSs. 
The 3FD results are compared with those of JAM \cite{Taya:2024zpv} of the cascade and mean-field (MF) versions,
where all matter (not necessarily equilibrated) was taken into account.  
}
\label{fig:V4-B-th_sNN_b=2fm_log}
\end{figure}

Analysis of four-volume of equilibrated baryonic matter, $V_4(\widetilde{n}_B^{\scr{equilibrated}})$, 
is advantageous because it focuses on effects of real compression of baryonic matter and 
excludes spurious kinematic effects (i.e. $\gamma$) and unphysical effects of the sharp-edge approximation. 
The advantage of the 3FD model is that it naturally treats the baryon stopping and equilibration. 
Therefore, the calculation of $V_4(\widetilde{n}_B^{\scr{equilibrated}})$ can be directly performed. 
In transport models, the analysis of equilibration is a separate and laborious procedure. 
Nevertheless, comparison of the 
3FD $V_4(\widetilde{n}_B^{\scr{equilibrated}})$ with JAM $V_4(\widetilde{n}_B^{\scr{all}})$
is still instructive. This comparison is presented in Fig. \ref{fig:V4-B-th_sNN_b=2fm_log}. 

Collision-energy dependence of four-volumes, in which the baryon density of the equilibrated matter 
exceeds values $\widetilde{n}_B/n_0=$ 3, 4 and 6 in the central ($b$ = 2 fm) Au+Au collisions
is presented in Fig. \ref{fig:V4-B-th_sNN_b=2fm_log}. These 3FD results are compared with four-volumes
calculated within cascade and mean-field (MF) versions of the JAM \cite{Taya:2024zpv}. 
For visual convenience, the areas between 3FD crossover and 1PT curves, as well as between JAM cascade and MF curves 
are shaded. Note that 
in the JAM calculations, four-volumes that are occupied by \textit{all} (i.e. not necessary equilibrated) matter
were evaluated. Nevertheless, the 3FD four-volumes noticeably exceed those in the JAM. 
This indicates that baryon stopping in the 3FD is stronger than that in JAM. 
This difference deserves a separate discussion, see sect. \ref{Discussion}. 

Leaving aside this difference for a while, let us turn to optimal collision-energy range for realizing macroscopic high baryon-density matter. Contrary to JAM, $V_4(3n_0)$ as a function of $\sqrt{s_{NN}}$ does not exhibit a maximum.  
It monotonously decreases with $\sqrt{s_{NN}}$ rise remaining larger than 900 fm$^4$/c (i.e. $\approx$ 5.5$^4$ fm$^4$/c). 
This is quite a macroscopic four-volume\footnote{
accordingly to the convention suggested in Ref. \cite{Taya:2024zpv}: the four-volume is called macroscopic 
if $V_4\gsim$  4$^4$ fm$^4$/c. 
This is just a definition that may or may not be appropriate depending on the specific application. 
}. 
$V_4(\widetilde{n}_B^{\scr{equilibrated}})$  exhibits  maxima in its dependence on $\sqrt{s_{NN}}$: 
maximal values $V_4(n_B^{\scr{equilibrated}}>4n_0)\gsim$  5$^4$ fm$^4$/c 
are reached at $\sqrt{s_{NN}}=$ 3.2--8 GeV, while 
$V_4(n_B^{\scr{equilibrated}}>6n_0)\gsim$  4$^4$ fm$^4$/c at $\sqrt{s_{NN}}=$ 4.5--9 GeV. 
Thus, even for $n_B/n_0>$ 6 the four-volume remains quite macroscopic, while it can hardly be said so within the 
JAM. The optimal energy range for densities $n_B/n_0>$ 4 is located approximately in the same region as 
that in the JAM (5--7 GeV) but the width of this range is considerably wider in the 3FD.

\section{Baryon Stopping vs EoS Stiffness}
\label{Discussion}

As mentioned above, the baryon stopping in the 3FD is stronger than in JAM. 
In fact, this difference in the baryon stopping correlates with stiffness of the EoS implemented in the 3FD and JAM 
models. The 3FD uses soft (in the hadronic phase) crossover and 1PT EoSs with incompressibility of 210 MeV
\cite{Toneev06}. This incompressibility was one of the constraints in construction of these crossover and 1PT EoSs
because the soft EoS is required for reproduction of the directed flow 
(more precisely, the directed transverse flow $\langle P_x/A \rangle$ \cite{Danielewicz:1985hn}) 
\cite{Danielewicz:2002pu}, as it was a common viewpoint that time. 
A stiff EoS produces a higher pressure in colliding nuclei, hence a stronger directed flow, which 
disagrees with data \cite{Danielewicz:1985hn}.
At the same time, 
the directed flow is insensitive to the stiffness of the EoS within the JAM 
\cite{Nara:2020ztb,Nara:2021fuu}. This looks counterintuitive, because the produced pressure 
turns out to be insensitive to the stiffness of the EoS.

On the other hand,  essential dependence on the EoS stiffness still remains in the UrQMD model \cite{Steinheimer:2022gqb}. 
The stiff EoS ($K\approx$ 360 MeV) is reported to be preferable at $\sqrt{s_{NN}}<$ 7 GeV. 
This definitely contradicts the conclusion of Ref. \cite{Danielewicz:1985hn}.
A natural question arises: what has happened since the analysis \cite{Danielewicz:1985hn}, 
in which the constraint on a soft EoS was deduced from data on the directed flow of protons?
The EoS stiffness and baryon stopping are two key factors in description of the directed flow.  
The pressure, which is required for the directed-flow formation, consists of two parts: the kinetic pressure, 
which results from the baryon stopping (i.e. randomization of nucleon momenta and production of secondary hadrons), 
and the potential pressure due to the EoS stiffness. If the kinetic pressure is high (strong baryon stopping) 
the lower potential pressure (a soft EoS) is needed to reproduce the directed flow. 
This is precisely the conclusion of the analysis in Ref. \cite{Danielewicz:1985hn}.
The 3FD supports this conclusion \cite{Ivanov:2025vru,Ivanov:2024gkn,Konchakovski:2014gda,Ivanov:2014ioa,Ivanov:2016sqy}. 
If the preference of stiff EoS is found \cite{Steinheimer:2022gqb}, that would mean that the 
baryon stopping in the UrQMD model \cite{Steinheimer:2022gqb} is weaker than in the analysis 
of Ref. \cite{Danielewicz:1985hn}  and in the 3FD model. 
Analysis of four-volumes (\ref{V4}) allows a direct comparison of 
the baryon stopping within different models. 
\\

\section{Summary}
\label{Summary}

Predictions of the 3FD model \cite{Ivanov:2005yw,Ivanov:2013wha} 
for the four-volumes (spatial 3-volume$\times$lifetime) are presented and compared with those of the JAM
\cite{Taya:2024zpv} in central Au+Au collisions at energies $\sqrt{s_{NN}}=$  3 -- 19.6 GeV. 
Analysis of these four-volumes (\ref{V4}) allows a direct comparison of 
the baryon stopping within different models.

It is found that the 3FD four-volumes noticeably exceed those in the JAM, which  
indicates a stronger baryon stopping in the 3FD model as compared to that JAM.
It is argued that this difference in the baryon stopping correlates with stiffness of the EoS 
implemented in the 3FD and JAM models.
The EoS stiffness and baryon stopping are two key factors in description of the directed flow.
The pressure, which is required for the directed-flow formation, consists of two parts: the kinetic pressure, 
that results from the baryon stopping (i.e. randomization of nucleon momenta and momenta of secondary hadrons), 
and the potential pressure due to the EoS stiffness. If the kinetic pressure is high (strong baryon stopping) 
a lower potential pressure (a soft EoS) is needed to reproduce the directed flow. 
This is the case in the 3FD model. 

Contrary to JAM, the four-volume, where a baryon density ($n_B$) exceeds 
three times the normal nuclear density ($n_0$), does not exhibit a maximum
as a function of $\sqrt{s_{NN}}$.  
It decreases monotonically with increasing $\sqrt{s_{NN}}$, remaining greater than 900 fm$^4$/s 
(i.e. $\approx$ 5.5$^4$ fm$^4$/c). 
This is quite a macroscopic four-volume accordingly to the convention suggested in Ref. \cite{Taya:2024zpv}: 
the four-volume is called macroscopic if $V_4\gsim$  4$^4$ fm$^4$/c. 
For higher baryon densities, $V_4$  exhibits  maxima in its dependence on $\sqrt{s_{NN}}$: 
for $n_B/n_0>$ 4, maximal values are reached at $\sqrt{s_{NN}}=$ 3.2--8 GeV, while 
for $n_B/n_0>$ 6,  at $\sqrt{s_{NN}}=$ 4.5--9 GeV. 
Even for $n_B/n_0>$ 6, the four-volume remains quite macroscopic ($V_4\gsim 4^4$ fm$^4$/c), 
while it is not so within the JAM. 
The optimal energy range for densities $n_B/n_0>$ 4 is located approximately in the same region as 
that in the JAM (5--7 GeV) but the width of this range is wider in the 3FD.

\begin{acknowledgments}

This work was carried out using computing resources of the federal collective usage center ``Complex for simulation and data processing for mega-science facilities'' at NRC "Kurchatov Institute" \cite{ckp.nrcki.ru}.

\end{acknowledgments}

\section*{DATA AVAILABILITY}

Tabulated 1PT and crossover EoSs that were used in the present simulations are publicly available 
on GitHub \cite{github.com/marinakozh}. 
The data that support the findings of this article are openly available.


\end{document}